\pdfoutput=1

\documentclass[11pt]{article}

\usepackage[]{emnlp2021}

\usepackage{times}
\usepackage{latexsym}

\usepackage[T1]{fontenc}

\usepackage[utf8]{inputenc}

\usepackage{microtype}

\usepackage{amssymb}
\usepackage{amsmath}
\usepackage{caption}
\usepackage{amsfonts}
\usepackage{romannum}
\usepackage{multirow}
\usepackage{graphicx}
\usepackage{booktabs}
\usepackage{subfigure}

\DeclareMathOperator*{\argmax}{arg\,max}

\title{Refining BERT Embeddings for Document Hashing via Mutual Information Maximization}

\author{
	Zijing Ou$^{1}$,~ Qinliang Su$^{1}$\thanks{~~Corresponding author. Qinliang Su is also affiliated with (\romannum{1}) Guangdong Key Lab. of Big Data Analysis and Processing, Guangzhou, China, and (\romannum{2}) Key Lab. of Machine Intelligence and Advanced Computing, Ministry of Education, China.},~ Jianxing Yu$^{2}$,~ Ruihui Zhao$^{3}$, \\
	\bf~ Yefeng Zheng$^{3}$, \and Bang Liu$^4$\\
	 $^1$School of Computer Science and Engineering, Sun Yat-sen University, Guangzhou, China, \\ 
	 $^2$School of Artificial Intelligence, Sun Yat-sen University, Guangdong, China, \\
	 $^3$Tencent Jarvis Lab, $^4$RALI \& Mila, Université de Montréal, \\
	ouzj@mail2.sysu.edu.cn, \{suqliang, yujx26\}@mail.sysu.edu.cn, \\ 
	\{zacharyzhao, yefengzheng\}@tencent.com, bang.liu@umontreal.ca \\
}

\begin{document}
\maketitle
\begin{abstract}
Existing unsupervised document hashing methods are mostly established on generative models. Due to the difficulties of capturing long dependency structures, these methods rarely model the raw documents directly, but instead to model the features extracted from them ({\it e.g.} bag-of-words (BOW), TFIDF). In this paper, we propose to learn hash codes from BERT embeddings after observing their tremendous successes on downstream tasks. As a first try, we modify existing generative hashing models to accommodate the BERT embeddings. However, little improvement is observed over the codes learned from  the old BOW or TFIDF features. We attribute this to the reconstruction requirement in the generative hashing, which will enforce irrelevant information that is abundant in the BERT embeddings also compressed into the codes. To remedy this issue, a new unsupervised hashing paradigm is further proposed based on the mutual information (MI) maximization principle. Specifically, the method first constructs appropriate global and local codes from the documents and then seeks to maximize their mutual information. Experimental results on three benchmark datasets demonstrate that the proposed method is able to generate hash codes that outperform existing ones learned from BOW features by a substantial margin.\footnote{Our code is available at \href{https://github.com/J-zin/DHIM}{https://github.com/J-zin/DHIM}.}

\end{abstract}

\section{Introduction}
With the explosion of information, similarity search \cite{jing2008visualrank} plays a increasingly important role in modern information retrieval systems. Traditional search engines conduct query by evaluating the distances of items in the \textit{continuous} Euclidean space, making it suffer from high computational complexity and footprint. To address this issue, considerable  efforts have been devoted to semantic hashing \cite{salakhutdinov2009semantic}, which aims to represent each document by a compact binary code. Such representations are able to reduce the memory footprint and increase the retrieval efficiency significantly by enrolling in \textit{binary} Hamming space.

A pivotal challenge in learning high-quality hash codes is how to retain the semantic similarities among documents. Although using supervised information is an efficient way to achieve this goal, due to the high cost of labeling, unsupervised hashing is more favourable in practice. Currently, most of  unsupervised document hashing methods are established upon the perspective of deep generative models \cite{kingma2013auto,rezende2014stochastic}. Essentially, all these methods seek to model the documents with a deep generative model and then employ the latent representations of documents to construct hash codes \cite{chaidaroon2017variational,shen2018nash,dong2019document,ye2020unsupervised,zheng2020generative,ou2021integrating}. Although great successes have been observed in these methods, due to the difficulties in capturing the long dependency structures of words (especially for long documents), all of these methods are established on modeling the BOW or TFIDF features of documents.

Although the BOW or TFIDF features are informative and are prevalent in many areas, their limitations are also obvious for not considering the word order and dependency structure. Recently, large-scale pre-trained language models like BERT \cite{devlin2018bert} have demonstrated their superior capabilities on various natural language understanding tasks. Embeddings extracted from them have also been shown to contain much more abundant information. Thus, in this paper, we argue that capitalizing on BERT embeddings to produce hash codes is better than on the out-of-date BOW features. As a first try, we modify existing generative hashing methods to accommodate the BERT embeddings and then use the trained model to generate hash codes. However, experimental results show no improvement on the quality of obtained hash codes. Even worse, the codes sometimes perform even inferior to those learned from BOW features. We conjecture that this is because  the reconstruction requirement in generative hashing enforces most of the information in BERT embeddings to be transferred into the hash codes. However, as the information contained in BERT embeddings is very abundant, with only a small proportion relevant to hashing, it is not surprising to see that the codes are not aligned well with the semantic similarities of documents.

To generate high-quality hash codes from BERT embeddings, it becomes necessary to refine the embeddings to highlight the information relevant to hashing tasks ({\it i.e.}, semantic information), while attenuating the irrelevant. Recent progresses on image representation learning have shown that it is possible to learn discriminative semantic representations using the mutual information (MI) maximization principle. Inspired by this, rather than utilizing the reconstruction structure, an alternative paradigm is proposed for unsupervised document hashing based on the MI maximization principle, named Deep Hash InfoMax (DHIM). The essential idea behind our approach is to construct appropriate global and local codes and then seek to maximize their mutual information, with the global and local codes accounting for the entire document and text fragments, respectively. As explained in image representation learning,  doing so implicitly encourages the global codes to retain high-level semantic information shared across different local fragments, while ignoring the local irrelevant details. Extensive experiments are conducted on three benchmark datasets. The results demonstrate that by effectively refining the BERT embeddings via MI maximization principle, the proposed method is able to generate hash codes that outperform existing ones learned from BOW features by a substantial margin.

\section{Preliminaries on Generative Hashing for Documents}
Document hashing aims to learn close binary codes for semantically similar documents. An intuitive idea towards this goal is to encourage hash codes preserving as much information of documents as possible so that close codes are easier to be obtained for similar documents. Based on this idea, many methods have be proposed to employ generative models like VAEs to model the documents and then leverage the documents' latent representations to produce binary hash codes. However, due to the difficulties in capturing the long dependency structures of words (especially for long documents), existing generative hashing methods rarely seek to model the documents directly, but instead to first extract representative features from documents ({\it e.g.}, BOW or TFIDF) and then perform modeling on the extracted features. Specifically, by representing a document $x$ as a sequence of words $x = \{w_1, w_2, \dots, w_{|x|}\}$, existing generative hashing methods \cite{chaidaroon2017variational} are mostly established on the following document model
\begin{align} \label{joint_prob_model}
	p(x,z) = \prod\limits_{w_i \in x} p_\theta (w_i | z) p(z),
\end{align}
where 
\begin{align} \label{bag_of_word_prob}
	p_\theta (w_i | z) \triangleq \frac{\operatorname{exp}(z^T E w_i + b_i )}{\sum_{j=1}^{|V|} \operatorname{exp}(z^T E w_j + b_j)}.
\end{align}
Here $z$ is the latent variable; $w_j$ is a $|V|$-dimensional one-hot vector corresponding to the $j$-th word; $E \in \mathbb{R}^{m \times |V|}$ represents the learnable embedding matrix; $b_i$ is the biased term; and $|V|$ and $|x|$ represent the vocabulary size and document length, respectively. The whole model is trained by maximizing the evidence lower bound (ELBO) of log-likelihood 
\begin{align} \label{elbo}
	{\mathcal{L}}(\theta, \phi) = {\mathbb{E}}_{q_\phi(z|x)}\left[\log \frac{p_\theta(x, z)}{q_\phi(z|x)} \right]
\end{align}
with respect to $\theta$ and $\phi$, where $q_\phi(z|x)$ denotes the approximate posterior distribution parameterized by $\phi$. After training, representation of the document $x$ can be extracted from the approximate posterior $q_\phi(z|x)$, {\it e.g.}, using its output mean. Note that a simple decoder of \eqref{bag_of_word_prob} is adopted purposely for better transferring similarity information of documents $x$ into the latent representations $z$.

In the early generative hashing work VDSH \cite{chaidaroon2017variational}, Gaussian distributions are employed for both the prior $p(z)$ and approximate posterior $q_\phi(z|x)$ directly. But due to the continuous characteristics of Gaussian random variables, a separate binarization step is required to transform the  continuous latent representations into binary codes. To overcome the separate training issue, Bernoulli prior and posterior are then proposed in NASH \cite{shen2018nash}.
With the recent advances on gradient estimators for discrete random variables, the model successfully circumvents gradient backpropagation issue for discrete variables, and can be trained efficiently in an end-to-end manner. Inspired by NASH, many variant methods are then proposed by using more sophisticated prior or posterior distributions, with the objective to model the documents more accurately, such as Bernoulli mixture prior in BMSH \cite{dong2019document} and Boltzmann machine posterior in CorrSH \cite{zheng2020generative} etc. 

Despite of the observed remarkable performance, all of the methods mentioned above rely on the document model \eqref{joint_prob_model}, which, however, is essentially established on the BOW features of documents, without considering any word order and dependency information. Although BOW features are informative, their limitations are also obvious due to the neglect of word order and dependency structure. With the development of large-scale pre-trained models like BERT, it becomes easy to obtain semantics-rich features that contain long dependencies and contextual information. Thus, we argue that it is beneficial to capitalize on the information-rich BERT embeddings over the out-of-date BOW features to learn hash codes.

\section{Hashing on BERT Embeddings via Generative Models} \label{Sec_GenerativeHashing}
Feeding a document $x = \{w_1, w_2, \cdots, w_{|x|}\}$ into a pre-trained BERT model could produce an embeding/feature for the document, denoted as ${\mathcal{B}}(x)$ for subsequent presentation. Inspired by the success of generative hashing methods, we modify them to accommodate the BERT embeddings. Due to the difference between  BERT embeddings and BOW features, the decoder in \eqref{bag_of_word_prob} is replaced by a conditional Gaussain distribution
\begin{align}
	p_\theta \left(x|z\right) = \frac{1}{(2\pi \sigma^2)^{d/2}} e^{-\frac{||\mathcal{B}(x) - W z ||^2}{2\sigma^2}  },
\end{align}
where $W$ is the learnable model parameter and the bias term is omitted for brevity; and $d$ denotes the dimension of BERT embeddings. Similar to the generative hashing models introduced above, here a simple decoder is employed purposely to facilitate the transferring of similarity information of BERT embeddings into the latent codes $z$. To achieve end-to-end training and directly output binary codes, Bernoulli prior $p(z)$ and approximate posterior $q_\phi(z|x)$ can be used, as done in NASH, BMSH etc. After training, the binary hash code of document $x$ can be obtained from the latent codes $z \sim q_\phi(z|x)$.

Unexpectedly, as observed in experiments (see Table \ref{table:results}), the codes generated from BERT embeddings in this manner perform even worse than that from TFIDF features. At first glance, this is unreasonable, since information in BERT embeddings is much abundant. However, we ought to emphasize that more information does not represent better performance. Although the BERT embedding has been successfully applied to various downstream tasks, it is also reported that directly using BERT embeddings can not yield satisfactory gains to information retrieval \cite{reimers2019sentence}. \citet{li2020sentence} attributed this issue to that the embedding contains many types of information, and the semantic information is not appropriately preserved. In this regard, the worse performance of naively exploiting BERT embeddings is traceable. In the generative hashing approach, what the model does basically is to compress the embedding ${\mathcal{B}}(x)$ into a latent code $z$ and then use the code to reconstruct the original embedding ${\mathcal{B}}(x)$. Due to the requirement of reconstruction, latent codes $z$ are enforced to preserve as much information of original inputs ${\mathcal{B}}(x)$ as possible. However, as discussed above, BERT embeddings contain various kinds of information, and the categorical information is just the one relevant to the hashing performance while the others are redundant. Thus, when the generative approach is applied to BERT embeddings, it is not surprising to see that the codes are not aligned well with the semantic similarities of documents.

\section{Refining BERT Embeddings via MI Maximization}

According to discussions in Section \ref{Sec_GenerativeHashing}, to produce high-quality hash codes, it is necessary to refine BERT embeddings to highlight the category-relevant information, while attenuating the other types of information. Recent progresses on image representation learning \cite{hjelm2019learning} have demonstrated that it is possible to learn category-discriminative representations from images unsupervisedly with the MI maximization principle. Inspired by this, a brand new hashing framework based on MI maximization principle is proposed, which learns binary hash cods from BERT embeddings without using the reconstruction requirement, thereby overcoming the issues associated with the generative hashing approaches.

\subsection{Deep InfoMax Review}
Deep InfoMax \cite{hjelm2019learning} learns category-discriminative representations for images by  maximizing the mutual information between global and local representations. It first constructs a global representation for an image and lots of local representations, both extracted from the image's CNN feature maps. Then, it estimates the MI between the global and local representations and  maximizes it. As explained in \cite{hjelm2019learning}, since there are many local representations and each of them accounts for a local region of an image, maximizing the global-local MI implicitly encourages the global representation to retain global semantic information that is shared across all local regions, while ignoring specific details exclusive to different local regions.

\subsection{Construction of Global/Local Document Features}
To refine BERT embeddings with the deep InfoMax, we first need to construct appropriate global and local document features. To this end, we re-represent a document as $X =\{e_1, \dots, e_{T}\}$, where $e_i \in {\mathbb{R}}^d$ is the BERT embedding of the $i$-th word in the document, and $T$ denotes the document length. Then, we pass the document $X$ through a textual CNN \cite{kim-2014-convolutional}, in which filters $W \in \mathbb{R}^{K \times n \times d}$ are convolved with the words sequentially, with $n$ and $K$ denoting the filter size and number, respectively. Obviously, such operation could generate local features for every piece of $n$-gram fragments. Specifically, the local feature for the $i$-th fragment is computed as
\begin{align}
    h_i^{(n)} = ReLU(W \ast e_{i:i+n-1}),
\end{align}
where $\ast$ denotes the convolution operator, and the bias term is omitted for brevity; and $ReLU(\cdot)$ represents the rectified linear unit (ReLU) function. By applying this filter to all text fragments, we obtain the local feature maps at all locations
\begin{align}
    H^{(n)} = \{h_1^{(n)}, h_2^{(n)}, \dots, h_T^{(n)}\}.
\end{align}
By passing $H^{(n)}$ to READOUT function, which can be a simple mean-over-time pooling operation \cite{collobert2011natural} or more sophisticated self-attention mechanism \cite{vaswani2017attention}, we obtain the document's global feature.

To further highlight the semantic information in global features, we propose to compute multi-granularity local and global features using different window sizes of convolution operation (set as $\{1,3,5\}$ in our experiments) . That is, the final local and global features are computed as
\begin{align}
	h_i &= \operatorname{MLP} (\operatorname{CONCAT}(\{h_i^{(n)}\}_{n \in \mathcal{N}})), \nonumber \\
	H &= \operatorname{READOUT}(\{h_i\}_{i=1}^T), \nonumber
\end{align}
where $\mathcal{N}$ denotes the set of different window sizes and MLP is the multilayer perception layer used to project the feature maps on desirable dimension. By maximizing the MI between global and local document features, the global feature $H$ is encouraged to keep high-level semantic information that are shared across all local fragments, while ignoring the irrelevant local details. 
\begin{figure}[!t]
	\centering
	\includegraphics[scale=0.53]{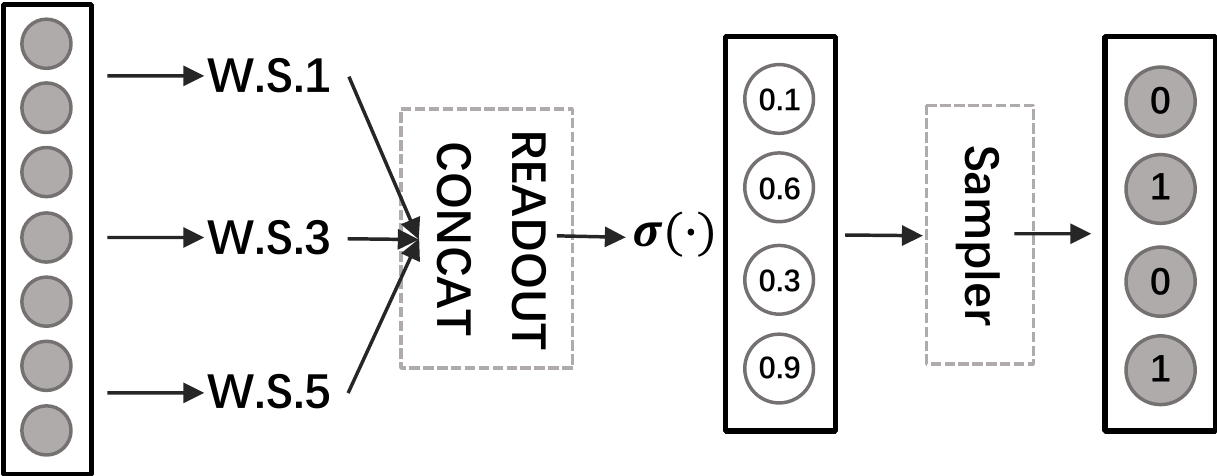}
	\caption{Architecture of the DHIM, in which W.S.n denotes convolution operation with window size n.}
	\label{architecture}
	\vspace{-2mm}
\end{figure}

\begin{figure*}[!t]
	\centering
	\includegraphics[scale=0.65]{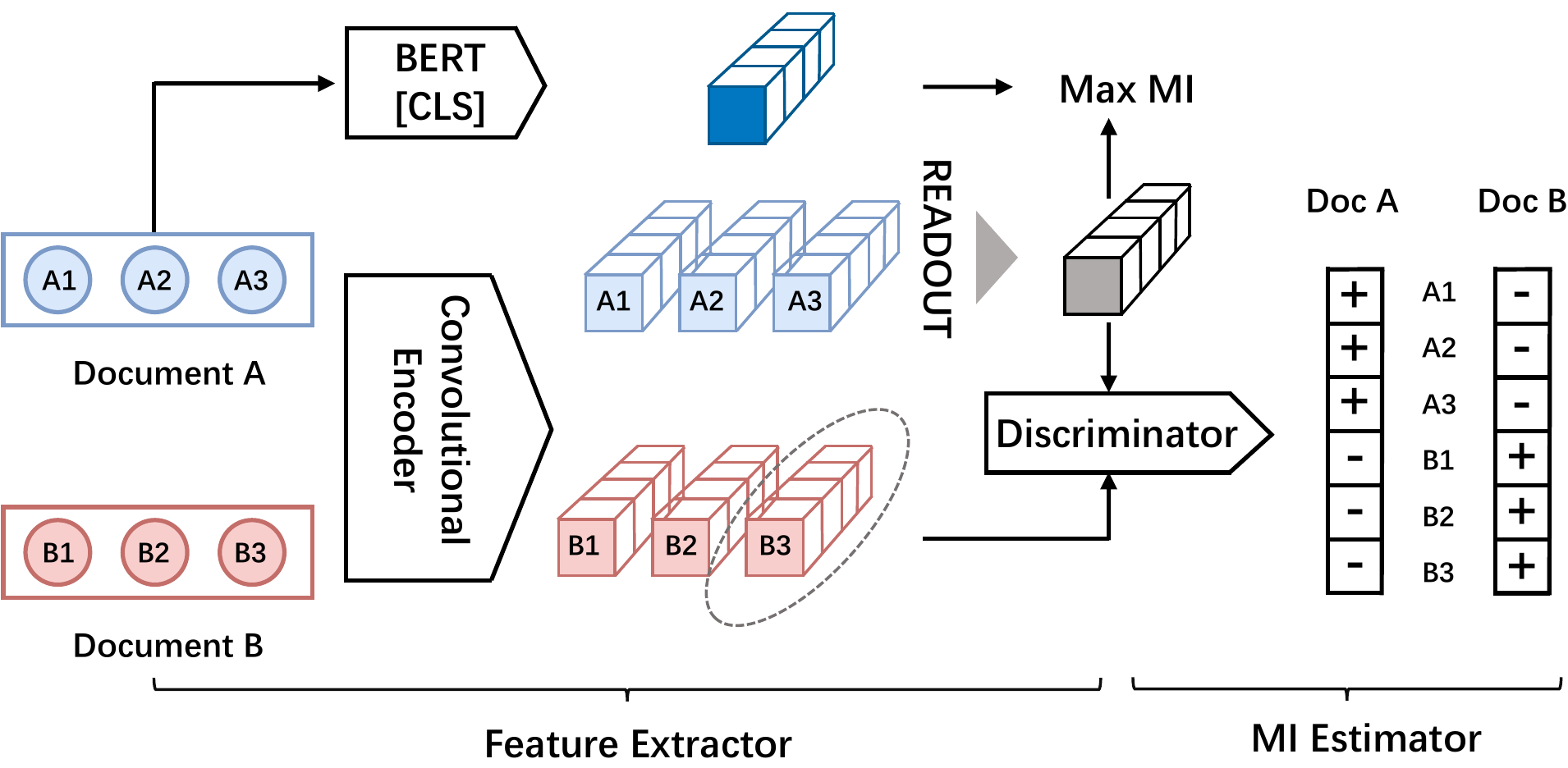}
	\caption{Intuitive illustration of DHIM. The local/global features are captured by textual convolution neural network and then fed into discriminator to identify whether they are from the same document. For example, consider a batch input with 2 documents with 3 words for each. For the global representation (gray cuboid) of document A, there will be 6 input pairs (local/global features) to the discriminator and same for document B. Additionally, we further encourage the mutual information between the learned representations and BERT CLS embedding to be high such that refining more semantic information into binary codes.}
	\label{image_objective}
	\vspace{-2mm}
\end{figure*}

\subsection{End-to-End Hashing by Maximizing the Global-Local MI}
Maximizing the global-local MI is able to yield semantic-rich global features $H$, which, however, are in the real-valued space. To obtain the binary hash codes, a feasible way is to binarize the global feature $H$, {\it e.g.}, by setting a threshold value. Obviously, the separate binarization strategy is not optimal in producing high-quality of codes. To obtain hashing models that admit end-to-end training, inspired by end-to-end generative hashing schemes, we propose to generate binary global and local representations by adding a probabilistic Bernoulli layer, that is,
\begin{align} \label{binarize_global_rep}
	b_i &\sim \operatorname{Bernoulli}(\sigma(h_i)), \nonumber \\
	B &\sim \operatorname{Bernoulli}(\sigma(H)) ,
\end{align}
where $b_i$ and $B$ denote local and global binary representations, respectively; and $\sigma(\cdot)$ denotes the sigmoid function that transforms the features into probability. The probabilistic binarization layer allows the gradient to be estimated efficiently by backpropagation-like algorithms like ST \cite{bengio2013estimating}, Gumbel-softmax \cite{jang2016categorical} etc., which are also widely used in the end-to-end generative hashing models. The overall architecture of generating binary representations is depicted in Figure \ref{architecture}.

Since our goal is to learn binary hash codes, instead of maximizing the MI between $H$ and $h_i$, we propose to maximize the MI between the global and local binary features $B$ and $b_i$ directly
\begin{align} \label{maximize_mi}
    \hat{\theta} = \argmax\limits_\theta \frac{1}{T}  \sum\limits_{i=1}^T\!\! I(b_i; B),
\end{align}
where $\theta$ is the model parameters involved in the construction of global and local binary representations. Note that $b_i$ and $B$ are not specific for one document, but for all documents in the training set. Mutual information is notoriously hard for evaluation. Recently, many sophisticated methods have been proposed to estimate it, such as MINE \cite{belghazi2018mutual},  infoNCE \cite{oord2018representation} and Jensen-Shannon divergence estimator (JSDE) \cite{nowozin2016f}. Among them, JSDE is known to be less sensitive to the number of negative samples, thus we apply it to estimate the MI and then optimize it w.r.t. the model paramters. Specifically, the MI can be estimated by minimizing the following function w.r.t. $\phi$
\begin{align} \label{MI_estimator}
    \tilde I_\phi (b_i; B) \!= &-\operatorname{softplus}(-D_{\phi}(b_i, B)) \nonumber \\ 
    &- \mathbb{E}_{\Tilde{\mathbb{P}}}[\operatorname{softplus}(D_\phi(\Tilde{b}_i, B))], 
\end{align}
where $\Tilde{b}_i$ is the $i$-th local representation of negative samples generated from empirical distribution $\Tilde{\mathbb{P}}= \mathbb{P}$; softplus function is defined as $\operatorname{softplus}(x) \triangleq \log (1 + e^x)$; and $D_{\phi}(\cdot, \cdot)$ is a discriminator realized by a neural network with parameter $\phi$. In practice, negative sample $\tilde b_i$ is chosen from local representations of other documents in a minibatch.

The MI  maximization scheme above relies solely on BERT embeddings of individual words, totally ignoring the embedding corresponding to the CLS token of BERT. The CLS embeddings are known to preserve global information of sentences or documents. Thus, to improve the global semantic information in the learned codes, we add a regularization term to boost the MI between the codes and CLS embedding. Therefore, the final loss takes the form
\begin{align}
     \mathcal{\tilde L}(\phi, \theta) \!=\! -\frac{1}{T} \! \sum\limits_{i=1}^T \tilde I_\phi(b_i ; B) \!-\! \beta \tilde I_\phi(E ; B),
\end{align}
where $\beta$ is a hyper-parameter; and $E$ denotes the binarized CLS embedding, obtained in similar way to \eqref{binarize_global_rep}. Note that $\theta$ is the model parameters involved in the construction of $b_i$ and $B$, while $\phi$ is used in the discriminator $D_\phi(\cdot, \cdot)$. By resorting to the gradient estimator for discrete random variables, the loss $\mathcal{\tilde L}(\phi, \theta)$ can be optimized efficiently with stochastic gradient decent (SGD) algorithms. An overall depiction of the proposed \textbf{D}eep \textbf{H}ash \textbf{I}nfo\textbf{M}ax (DHIM) model is illustrated in Figure \ref{image_objective}.

\section{Related Work}
Early works in unsupervised document hashing generally built upon the generative models \cite{kingma2013auto,rezende2014stochastic}, in which the encoder-decoder architecture was established to encourage binary codes to retain semantic information by reconstructing original data. For examples, VDSH \cite{chaidaroon2017variational} first proposed to learn continuous representations under variational autoencoder (VAE) framework, and then cast it into binary codes. However, the two-stage training procedure is prone to undermine the performance. NASH \cite{shen2018nash} tackled this issue by replacing Gaussian prior with Bernoulli in VAE and adopting straight-through to enable end-to-end training. Since then, a lot of methods surged to improve the performance. 

Specifically, \citet{dong2019document} proposed to employ mixture distribution as prior to enhance model's capabilities; \citet{ye2020unsupervised} introduced auxiliary topic vectors to address the problem of information loss in few-bits scenarios, and \citet{zheng2020generative} employed Boltzmann posterior to introduce correlation among bits. Beyond generative models, AMMI \cite{stratos2020learning} achieved superior performance by maximizing mutual information between documents and codes. However, the adversarial training procedure used in AMMI is extremely unstable. Although these models are impressive, one common issue of them is that they simply exploited bag-of-words features as input, which is not enough to capture the rich semantic information of documents.

Recently, information theory enables a simple and insightful paradigm of unsupervised representation learning \cite{oord2018representation,stratos2020learning,qiu2021unsupervised}. For example, \citet{hjelm2019learning} proposed an unsupervised representation learning algorithm on image data, called Deep InfoMax, which maximizes the MI between the whole image and local patches. \citet{velickovic2019deep} and \citet{sun2019infograph} extended this idea on graph data, in which the representations can be learned by maximizing the MI between the whole and sub graphs. These methods consistently encourage the global representations to retain similar interest of local features. Following similar ideas, we train our models that maximize MI between local n-grams features and the pooled global document representation, which can efficiently distill the semantic information of BERT embedding into hash codes.

\section{Experiments}
\subsection{Experiment Setup}

\paragraph{Datasets} We verify the proposed model on three public benchmark datasets: \romannum{1}) The New York Times (NYT) \cite{tao2018doc2cube}, which contains news articles published by The New York Times; \romannum{2}) DBpedia \cite{lehmann2015dbpedia}, which contains the abstract of articles extracted from Wikipedia; \romannum{3}) AGNews \cite{zhang2015character}, which is a news collection gathered from academic news search engine. For all documents in a dataset, we simply apply the same string cleaning operation\footnote{\href{https://github.com/yoonkim/CNN\_sentence/blob/master/process_data.py}{https://github.com/yoonkim/CNN\_sentence}} conducted in \cite{kim-2014-convolutional}. After that, it is randomly split into training, validation and test sets, with the statistics shown in Table \ref{table:statistic}.

\begin{table}[!t]
\centering
\setlength{\tabcolsep}{.7mm}{
    \begin{tabular}{cccccc}
    \toprule
    \textbf{Dataset} &  \textbf{Train}  &  \textbf{Val}  &  \textbf{Test} &  \textbf{Classes} & \textbf{AvgLen} \\
    \midrule
    NYT & 9,221 & 1,154 & 1,152 & 26 & 648 \\
    DBpedia & 50,000 & 5,000 & 5,000 & 14 & 47 \\
    AGNews & 114,839 & 6,381 & 6,380 & 4 & 32 \\
    \bottomrule
    \end{tabular}}
\caption{The statistic of three benchmark datasets.}
\vspace{-4mm}
\label{table:statistic}
\end{table}

\begin{table*}[!t]
	\centering
	\small
	\setlength{\tabcolsep}{1.2mm}{
		\begin{tabular}{c|cccc|cccc|cccc}
		\toprule
		\multirow{2}*{\textbf{Method}} & \multicolumn{4}{c}{NYT} & \multicolumn{4}{|c}{DBpedia} & \multicolumn{4}{|c}{AGNews}   \\
		\cmidrule(r){2-5} \cmidrule(r){6-9} \cmidrule(r){10-13}
		 & 16bits & 32bits & 64bits & 128bits & 16bits & 32bits & 64bits & 128bits & 16bits & 32bits & 64bits & 128bits  \\
		\midrule
		$\text{VDSH}^{\clubsuit}$ & 0.6877 & 0.6877 & 0.7501 & 0.7849 & 0.6779 & 0.7264 & 0.7884 & 0.8491 & 0.6732 & 0.6742 & 0.7270 & 0.7386  \\
		$\text{NASH}^{\clubsuit}$ & 0.7487 & 0.7552 & 0.7508 & 0.7301 & 0.7802 & 0.7984 & 0.7979 & 0.7676 & 0.6574 & 0.6934 & 0.7272 & 0.7433  \\
		$\text{WISH}^{\clubsuit}$ & 0.7015 & 0.7003 & 0.6448 & 0.6894 & 0.8228 & 0.8276 & 0.8210 & 0.7822 & 0.7453 & 0.7479 & 0.7505 & 0.7270  \\
		$\text{BMSH}^{\clubsuit}$ & 0.7402 & 0.7638 & 0.7688 & 0.7763 & 0.8317 & 0.8624 & 0.8705 & 0.8386 & 0.7409 & 0.7603 & 0.7609 & 0.7356  \\
		$\text{CorrSH}^{\clubsuit}$ & 0.7543 & 0.7761 & 0.7724 & 0.7839 & 0.8201 & 0.8178 & 0.8094 & 0.8577 & 0.7620 & 0.7645 & 0.7661 & 0.7767  \\
		$\text{AMMI}^{\clubsuit}$ & 0.7106 & 0.7648 & 0.7737 & 0.7803 & 0.8451 & 0.8953 & 0.9078 & \textbf{0.9103} & 0.7647 & 0.7661 & 0.7732 & 0.7823  \\
		\midrule
		$\text{VDSH}^{\spadesuit}$ & 0.5338 & 0.5818 & 0.6244 & 0.6464 & 0.6959 & 0.7521 & 0.7954 & 0.8062 & 0.6297 & 0.6635 & 0.6957 & 0.7027  \\
		$\text{NASH}^{\spadesuit}$ & 0.5587 & 0.5825 & 0.6098 & 0.6427 & 0.6587 & 0.7454 & 0.7796 & 0.8143 & 0.6632 & 0.6844 & 0.7040 & 0.7207  \\
		$\text{WISH}^{\spadesuit}$ & 0.5883 & 0.6475 & 0.6547 & 0.7034 & 0.6565 & 0.7291 & 0.7666 & 0.8229 & 0.6535 & 0.6619 & 0.6939 & 0.7203 \\
		$\text{BMSH}^{\spadesuit}$ & 0.5935 & 0.6326 & 0.6587 & 0.6971 & 0.6642 & 0.7913 & 0.8201 & 0.8457 & 0.6677 & 0.6961 & 0.7199 & 0.7316 \\
		$\text{CorrSH}^{\spadesuit}$ & 0.6203 & 0.6548 & 0.6838 & 0.7228 & 0.6528 & 0.7463 & 0.7865 & 0.8361 & 0.6706 & 0.6851 & 0.7086 & 0.7317 \\
		$\text{AMMI}^{\spadesuit}$ & 0.6047 & 0.6510 & 0.6967 & 0.7447 & 0.8025 & 0.8267 & 0.8926 & 0.8674 & 0.6550 & 0.6826 & 0.7185 & 0.7436 \\
		\midrule
		\textbf{DHIM} & \textbf{0.7969} & \textbf{0.8055} & \textbf{0.7977} & \textbf{0.7909} & \textbf{0.9426} & \textbf{0.9480} & \textbf{0.9302} & {0.8821} &  \textbf{0.7823} & \textbf{0.7917} & \textbf{0.7888} & \textbf{0.7986} \\
		\bottomrule
	\end{tabular}}
	\caption{The precision on three datasets with different numbers of bits in unsupervised document hashing. $\clubsuit$ and $\spadesuit$ denote that the input document features are TFIDF and BERT embeddings, respectively.}
	\vspace{-4mm}
	\label{table:results}
\end{table*}

\paragraph{Baselines} 
We compare our model with the following unsupervised deep semantic hashing methods: VDSH \cite{chaidaroon2017variational}, NASH \cite{shen2018nash}, BMSH \cite{dong2019document}, WISH \cite{ye2020unsupervised}, CorrSH \cite{zheng2020generative} and AMMI \cite{stratos2020learning}. The TFIDF features and BERT embeddings are taken as input to evaluate their impact for baselines. We exploit sklearn TfidfVectorizer API to extract TFIDF features for each document with the number of dimension in $10,000$, $20,000$, and $20,000$ for NYT, DBpedia and AGnews, respectively. BERT embedding is the CLS embedding, whose dimension is $768$. For all baselines, we tune their parameters on the validation set and select the best one to evaluate on the test set.

\begin{table}[!t]
\centering
\small
\setlength{\tabcolsep}{1.2mm}{
    \begin{tabular}{cc|ccccc}
    \toprule
    \multicolumn{2}{c|}{\textbf{Ablation Study}} &  16bits  &  32bits  &  64bits &  128bits \\
    \midrule
    \multirow{3}*{NYT}
    &  $\text{DHIM}_{\text{median}}$ &  0.7040   &  0.6949   &  0.6943 &  0.6999 \\
    &  $\text{DHIM}_{\text{w/o reg}}$  &  0.7371   &  0.7639   &  0.7704 &  0.7647 \\
    &  \textbf{DHIM}    &  \textbf{0.7969}   &  \textbf{0.8055}   &  \textbf{0.7977} &  \textbf{0.7909} \\
    \midrule
    \multirow{3}*{DBpedia}
    &  $\text{DHIM}_{\text{median}}$ &  0.7955   &  0.8432   &  0.8530 &  0.8630 \\
    &  $\text{DHIM}_{\text{w/o reg}}$ &  0.9057  &  0.9327  &  0.9206  &  0.8788  \\
    &  \textbf{DHIM} &  \textbf{0.9426}   &  \textbf{0.9480}   &  \textbf{0.9302} &  \textbf{0.8821} \\
    \midrule
    \multirow{3}*{AGnews}
    &  $\text{DHIM}_{\text{median}}$ &  0.7431   &  0.7538   &  0.7767 &  0.7897 \\
    &  $\text{DHIM}_{\text{w/o reg}}$  &  0.7629  &  0.7622  &  0.7821  &  0.7944  \\
    &  \textbf{DHIM} &  \textbf{0.7823}   &  \textbf{0.7917}   &  \textbf{0.7888} &  \textbf{0.7986} \\
    \bottomrule
    \end{tabular}}
\caption{The performance of variant models of DHIM.}
\vspace{-5mm}
\label{table:ablation_study}
\end{table}

\paragraph{Training Details} 
We implement our model with PyTorch and HuggingFace API \cite{DBLP:journals/corr/abs-1910-03771}. In our experiment, the discriminator $D_\phi$ is constituted by a one-layer feed-forward neural network followed with a sigmoid activation function, and the READOUT function is simply implemented as mean-pooling. We exploit the output of BERT-base module \cite{devlin2018bert}  as the features of documents. During training, the parameters of pre-trained BERT network are fixed, while only training the proposed convolutional encoder. We employ Adam optimizer for optimization \cite{kingma2014adam}, with the learning rate selected from $\{1 \times 10^{-3}, 1\times 10^{-4}, 1\times 10^{-5}\}$, and coefficient $\beta$ from $\{0.1, 0.2, \dots, 1\}$, according to the performance observed on the validation set.

\paragraph{Evaluation Metrics} 
Same as the previous works \cite{chaidaroon2017variational}, the retrieval precision is used to measure the quality of generated hash codes. For each query document, we retrieval its top-$100$ most similar documents based on the Hamming distance of learned codes. Then the retrieval precision is calculated as the percentage of the retrieved documents sharing with the same label as the query. Finally, The precision averaged over the whole test set is reported as the performance of the evaluated method.

\subsection{Results and Analysis}

\paragraph{Overall Performance}
The performances of our proposed model DHIM and all baselines are demonstrated in Table \ref{table:results}. It can be seen that our model performs favorably to the current state-of-the-art methods, yielding best performance across different datasets and settings. Compared with taking TFIDF as input, we find that the performance declines sharply if directly taking BERT embedding as input and redefining the generative model as Gaussian. This may be attributed to the fact that the reconstruction-based models may potentially tend to pay more attention on the generation of semantically-irrelevant information. However, if further refining the BERT embeddings via the proposed DHIM model, significant performance gains can be observed, which strongly corroborates the benefit of mutual information maximization framework. When examining the performance across different code lengths, our proposed method can achieve comparable performance with short codes. This is an attractive nature, since remarkable gratuity can be acquired profitably on the short codes, which is more suitable for low resource (small footprint) scenarios.
\begin{table}[!t]
\centering
\small
\setlength{\tabcolsep}{1.7mm}{
    \begin{tabular}{c|ccccc}
    \toprule
    \textbf{Features} &  16bits  &  32bits  &  64bits &  128bits \\
    \midrule
    Random  &  0.8140 & 0.8377 & 0.8666 &  0.8612 \\
    GloVe  &  0.8334 & 0.8507  &  0.8734 &  0.8611 \\
    $\text{BERT}_{\text{base}}$  &  {0.9426} & {0.9480} & \textbf{0.9302} & {0.8821} \\
    $\text{BERT}_{\text{large}}$  &  {0.9167} & {0.9261} & {0.9013} & \textbf{0.8902} \\
    $\text{ROBERTA}_{\text{base}}$  &  {0.9383} & {0.9437} & {0.9142} & {0.8728} \\
    $\text{ROBERTA}_{\text{large}}$  &  \textbf{0.9521} & \textbf{0.9527} & {0.9144} & {0.8706} \\
    \bottomrule
    \end{tabular}}
\caption{The performance of models with variant document features on the DBpedia datasets.}
\vspace{-5mm}
\label{table:diff_features}
\end{table}

\begin{table*}[!t]
\centering
\small
\setlength{\tabcolsep}{1.2mm}{
    \begin{tabular}{c|c|c}
    \toprule
    \textbf{Distance} & \textbf{Category} & \textbf{Content} \\
    \midrule
    \textbf{query} & \textbf{Athlete} & \textbf{Ilya Aleksandrovich Borodin (born July 6 1976) is a Russian professional footballer} \\
    1 & Athlete & Vojislav vodka Meli (5 January 1940 - 7 April 2006) is a former Yugoslavian footballer \\
    5 & Athlete & Rik Goyito Gregorio pérez (born November 19 1989) is a Mexican mixed martial artist \\
    10 & Artist & Themistocles Popa (June 27 1921 - November 26 2013) was a Romanian composer musician \\
    20 & Film & Allpakallpais a 1975 Peruvian drama film directed by Bernardo Arias \\
    30 & Transportation & USS Alcor (ad 34) was a destroyer tender the lone ship in her class named \\
    \bottomrule
    \end{tabular}}
\caption{Qualitative analysis of the learned 32-bit hash codes on the DBpedia dataset. We present the documents with Hamming distance of 1, 5, 10, 20 and 30 to the query.}
\label{table:case_study}
\vspace{-5mm}
\end{table*}

\paragraph{Ablation Study} To understand the influence of different components of DHIM, we further experiment with two variants of our model: \romannum{1}) $\text{DHIM}_{\text{median}}$: DHIM with hash codes after directly binarizing the real-value representations using the median value as the threshold; \romannum{2}) $\text{DHIM}_{\text{w/o reg}}$: DHIM without semantic-preserving regularizer. As seen from Table \ref{table:ablation_study}, $\text{DHIM}_{\text{w/o reg}}$ achieves better performance than $\text{DHIM}_{\text{median}}$, demonstrating the effectiveness of our proposed adaptions on the original deep InfoMax framework, {\it i.e.}, introducing a probabilistic layer to enable end-to-end training. Moreover, the additional semantic-preserving regularization is benefit to integrate expressive semantic information. This can be verified by significant performance of DHIM over $\text{DHIM}_{\text{w/o reg}}$, especially in short bits scenarios. However, the performance gap between them becomes small as code length increases. We attribute this interesting observation to the fact that the increased generalization ability of models brought by large bits is inclined to alleviate the impact of semantic regularization.

\begin{figure}[!t]
    \centering
	\subfigure{
		\begin{minipage}{0.45\columnwidth}
  	 	\includegraphics[width = 1.\columnwidth]{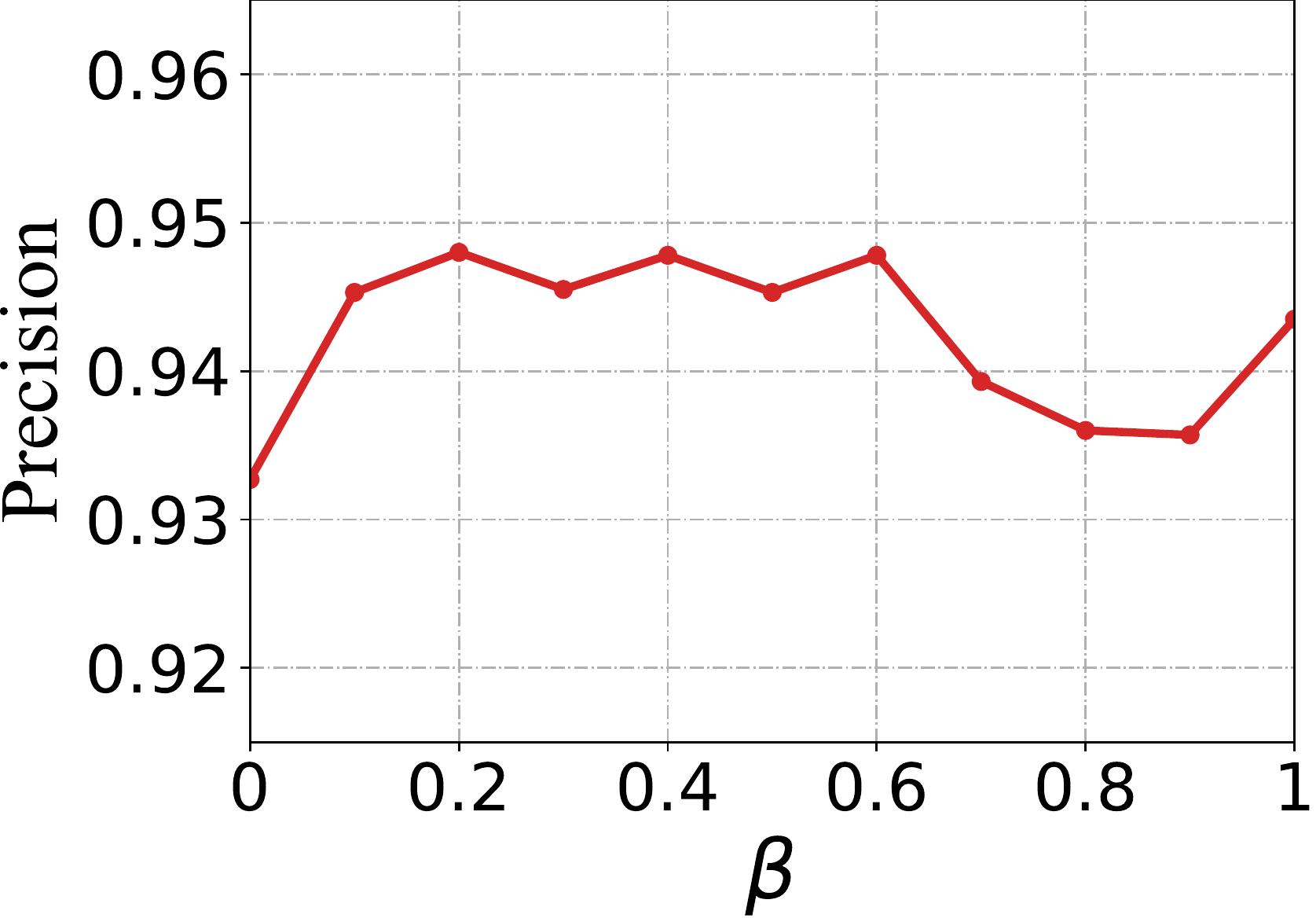}
		\end{minipage}
	}
	\subfigure{
		\begin{minipage}{0.45\columnwidth}
  	 	\includegraphics[width = 1.\columnwidth]{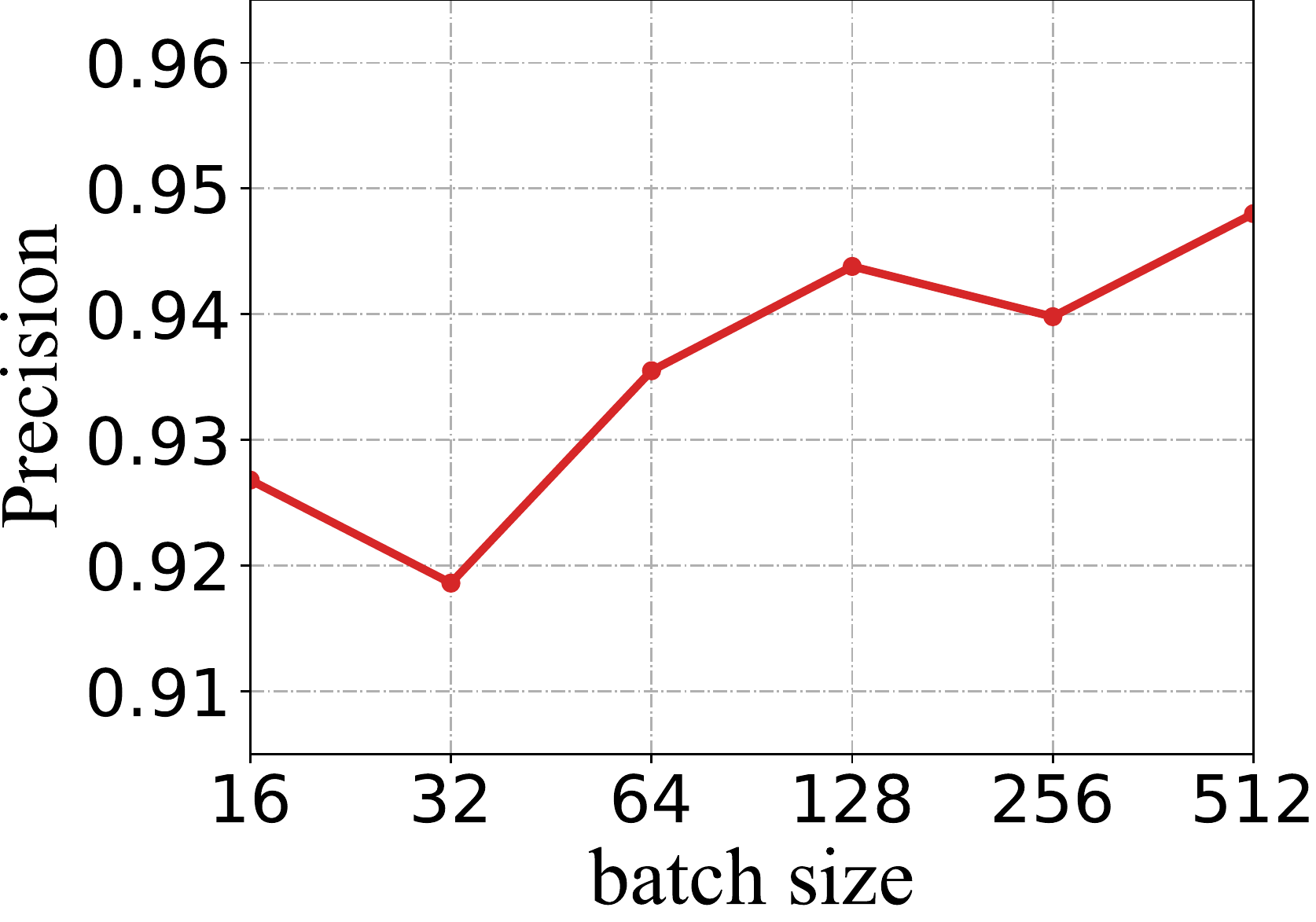}
		\end{minipage}
	}
	\vspace{-2mm}
	\caption{Parameter sensitivity analysis for the $\beta$ and batch size with 32-bit hash codes on DBpedia.}
	\vspace{-5mm}
	\label{fig: hypersensitivity}
\end{figure}

\paragraph{Impact of Different Features} One desirable property of DHIM is that we can exploit different textual features to enhance model abilities. To understand their effects, we investigate the impact of different kinds of word features: \romannum{1}) {Random}: with randomly initialized word embeddings; \romannum{2}) {GloVe}: with the GloVe embeddings \cite{pennington2014glove}; \romannum{3}) {Pre-trained}: with the ouputs of BERT \cite{devlin2018bert} or ROBERTA \cite{liu2019roberta}. As seen from Table \ref{table:diff_features}, simply exploiting random embeddings, our model still achieves comparable performance, demonstrating the effectiveness of the proposed mutual information maximization based hashing framework. It is worth to note that the model trained on pre-trained features yields better performance. This proves that the expressive context information of the document is conducive to learning high-quality hash codes.

\paragraph{Parameter Sensitivity} We also investigate the influence of hyperparameter $\beta$ and minibatch size. As shown in the left column of Figure \ref{fig: hypersensitivity}, compared with the case of $\beta = 0$, significant performance gains can be obtained by introducing semantic regularization. However, the appropriate value of $ \beta $ should be chosen carefully, since the best performance cannot be guaranteed if $\beta$ is too small or too large. Since the number of negative samples plays important roles in MI estimation, we further investigate the impact of batch size. From the right column of Figure \ref{fig: hypersensitivity}, we see that as batch size increases, the performance rises gradually and then converges to certain level.

\begin{figure}[!t]
    \vspace{-2mm}
	\centering
	\subfigure[DHIM]{
		\begin{minipage}[t]{0.4\linewidth}
			\centering
			\includegraphics[width=1.1in]{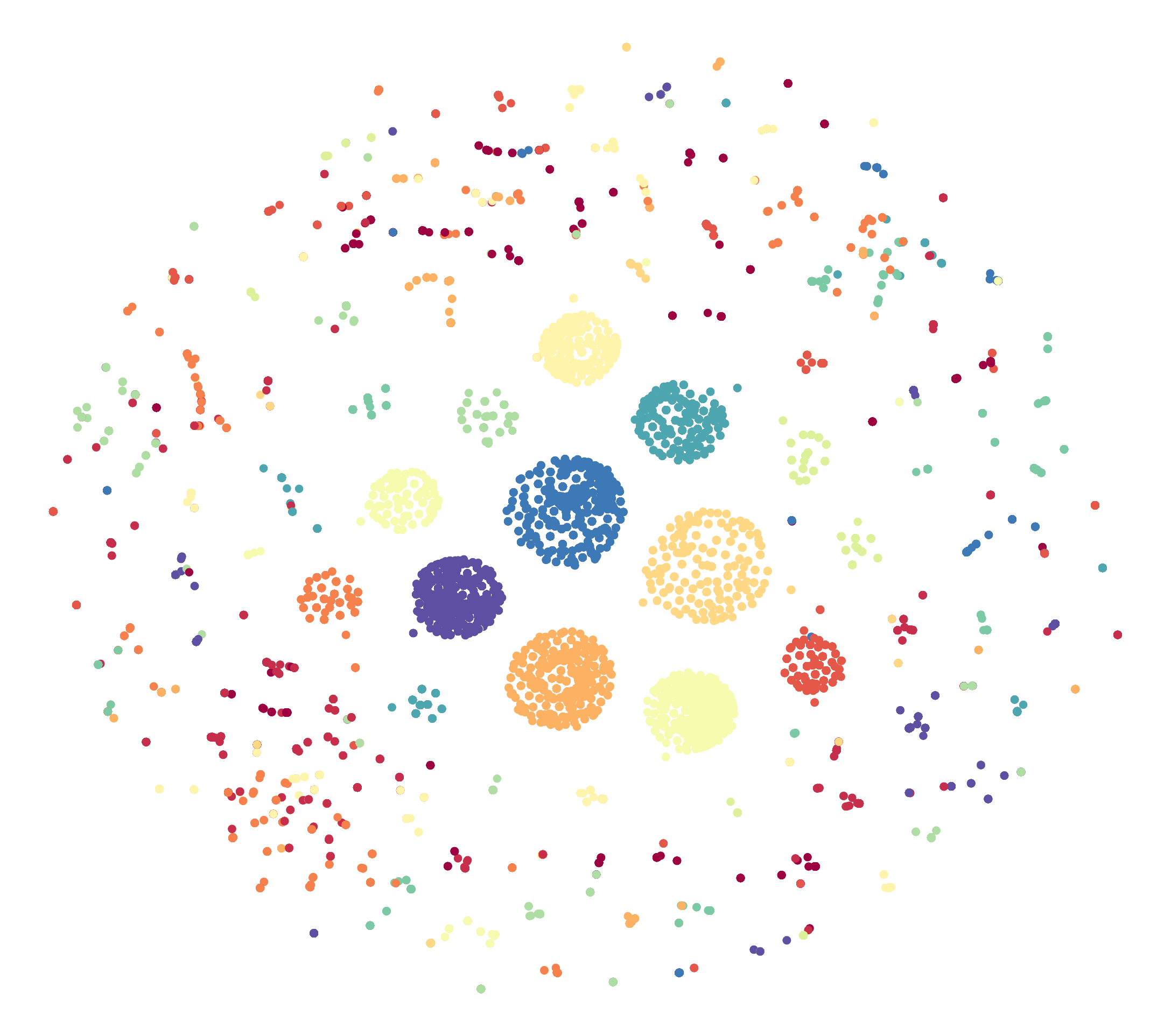}
		\end{minipage}
	}
	\subfigure[AMMI]{
		\begin{minipage}[t]{0.5\linewidth}
			\centering
			\includegraphics[width=1.3in]{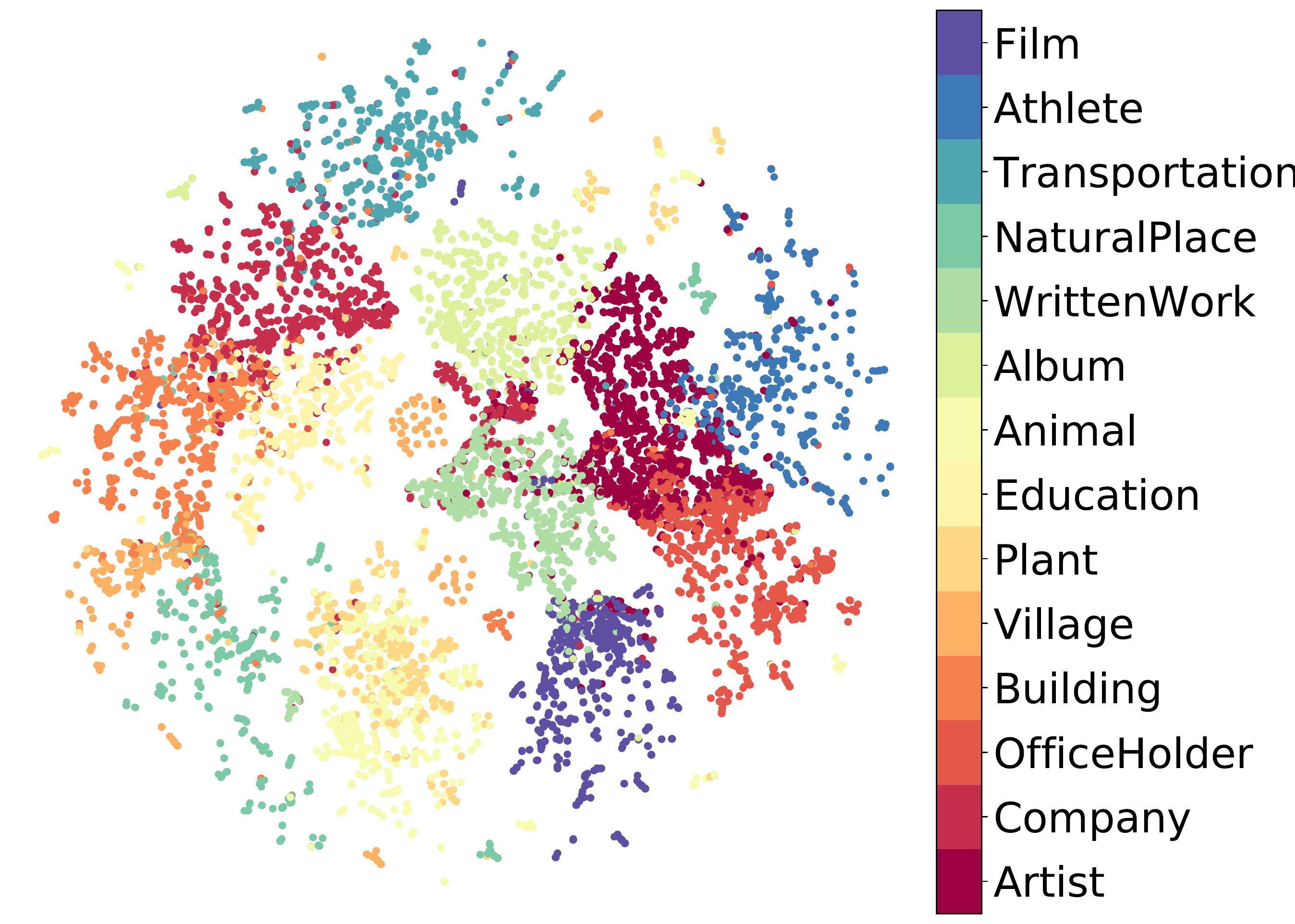}
		\end{minipage}
	}
	\vspace{-2mm}
	\caption{Visualization of the 32-bit codes learned by the proposed models for the DBpedia dataset.}
	\vspace{-5mm}
	\label{fig:hash_codes_visulization}
\end{figure}

\paragraph{Case Study}
To evaluate the quality of generated codes more intuitively, we present a retrieval case of the given query documents. As shown in Table \ref{table:case_study}, as the Hamming distance increases, the semantic of the retrieved document becomes less relevant, illustrating that the hash codes can effectively capture the semantic information.

\paragraph{Visualization of Hash Codes}
In Figure \ref{fig:hash_codes_visulization}, we project the learned binary codes into $2$-dimensional plane with t-SNE \cite{van2008visualizing} technique. It can be seen that the codes produced by our DHIM are more distinguishable than those of AMMI, demonstrating the superiority of our method.

\section{Conclusion}
We have proposed an effective and efficient semantic hashing method by refining the BERT embedding. Specifically, we applied a textual convolutional neural network with probabilistic layers to capture local and global features, and refined semantic information into binary codes by maximizing their mutual information. Extensive evaluations demonstrated that our model significantly outperforms baseline methods by learning hash codes under the guidance of MMI frameworks.

\section{Acknowledgement}
This work is supported by the National Natural Science Foundation of China (No. 61806223, U1811264), Key R\&D Program of Guangdong Province (No. 2018B010107005), National Natural Science Foundation of Guangdong Province (No. 2021A1515012299), Science and Technology Program of Guangzhou (No. 202102021205). This work is also supported by Huawei MindSpore.

\bibliography{custom}
\bibliographystyle{acl_natbib}

\end{document}